# Graphene oxide for high performance nonlinear optics in integrated nanowires

David Moss

Optical Sciences Centre, Swinburne University of Technology, Hawthorn, VIC 3122, Australia

## ABSTRACT

We report enhanced nonlinear optics in integrated nanophotonic chips through the use of integrated with 2D graphene oxide (GO) films. We investigate nanophotonic platforms including silicon, silicon nitride and high index doped silica. Due to the high Kerr nonlinearity of GO films and low nonlinear absorption we observe significant enhancement of third-order nonlinear processes. In particular, in silicon we observe an increase in both the Kerr nonlinearity and nonlinear figure of merit of up to 20 times. These results show the strong capability of GO films for improving the nonlinear optical performance of integrated photonic devices.

**Keywords:** nonlinear optics, self-phase modulation, parametric gain, silicon nitride waveguides, graphene oxide, Kerr nonlinearity

## 1. INTRODUCTION

All-optical signal generation, amplification, and processing based on optical nonlinearities offers processing speeds that far exceed that of purely electronic devices [1-3], for many applications including broadband optical amplification [4, 5], ultrafast switching [6, 7], optical logic gates [8, 9], all-optical wavelength conversion [10, 11], metrology [12, 13], spectroscopy [14, 15], optical cloaking [16, 17], and quantum optics [18, 19]. Compared to bulky discrete off-chip devices, photonic integrated circuits fabricated by well-established complementary metal-oxide semiconductor (CMOS) technologies provide an attractive solution to implement compact nonlinear optical devices on a chip scale, thus harvesting great dividends for integrated devices such as high stability and scalability, low power consumption, and large-scale manufacturing [20-22].

Although silicon-on-insulator (SOI) has been the dominant platform for photonic integrated circuits, its indirect bandgap is a significant handicap for optical sources, and its centrosymmetric crystal structure poses an intrinsic limitation for second-order nonlinear optical applications. Furthermore, its strong two-photon absorption (TPA) at near-infrared wavelengths limits its third-order nonlinear optical response in the telecom band [2, 23]. Other CMOS compatible platforms such as silicon nitride [7, 24, 25] and doped silica [26, 27] have a much lower TPA, although they still face the limitation of having a much smaller third-order optical nonlinearity than silicon. To address these issues, the on-chip integration of novel materials has opened up promising avenues to overcome the limitations of these existing integrated platforms. Many hybrid nonlinear integrated photonic devices incorporating polymers [28, 29], carbon nanotubes [30, 31], and two-dimensional (2D) materials [32-34] have been reported, showing significantly improved performance and offering new capabilities beyond those of conventional integrated photonic devices.

2D materials, such as graphene, black phosphorus (BP), transition metal dichalcogenides (TMDCs), hexagonal boron nitride (hBN), and graphene oxide (GO), have motivated a huge upsurge in activity since the discovery of graphene in 2004 [35]. With atomically thin and layered structures, they have exhibited many remarkable optical properties that are intrinsically different from those of conventional bulk materials [36-42]. Recently, there has been increasing interest in the nonlinear optical properties of 2D materials, which are not only fascinating in terms of laboratory research but also intriguing for potential practical and industrial applications [43-50].

Amongst the different 2D materials, GO has shown many advantages for implementing hybrid integrated photonic devices with superior nonlinear optical performance [37, 51-54]. It has been reported that GO has a large third-order optical nonlinearity ($n_2$) that is over 4 orders of magnitude higher than silicon [55, 56] as well as a linear absorption that is over 2 orders of magnitude lower than graphene in the infrared region [52, 57], both of which are very useful for third-order nonlinear optical processes. In addition, GO has a heterogenous atomic structure that exhibits non-centrosymmetry, yielding a large second-order optical nonlinearity that is absent in pristine graphene that has a centrosymmetric structure. The bandgap and defects in GO can also be engineered to facilitate diverse linear and nonlinear optical processes. These material properties of GO, together with its facile synthesis processes and high compatibility with integrated platforms [57, 58], have enabled a series of high-performance nonlinear integrated photonic devices. Here, we review our work on nonlinear integrated photonics based on the integration of 2D graphene oxide films, highlighting their applications based on a range of nonlinear optical processes (Figure 1) as well as a comparison of the different integrated platforms.

As an important third-order nonlinear optical process arising from Kerr effect, self-phase modulation (SPM) that occurs when an optical pulse with a high peak power propagates through a nonlinear medium has been widely utilized for broadband optical sources, pulse compression, optical spectroscopy, and optical coherence tomography [59-62 1-4]. The ability to realize SPM based on-chip integrated photonic devices will reap attractive benefits of compact footprint, high stability, high scalability, and low-cost mass production [63- 66  5-8]. Although silicon has shown itself to be a dominant platform for integrated photonic devices [61 3], it suffers from strong two-photon absorption (TPA) at near-infrared wavelengths, which greatly limits the nonlinear performance, and this has motivated the use of highly nonlinear materials on chips. Other complementary metal-oxide-semiconductor (CMOS) compatible platforms including high index doped silica glass (Hydex) [63 5] and silicon nitride ($Si_3N_4$) [62 4] have a much lower TPA, but they hamper the nonlinear performance due to a comparatively low Kerr nonlinearity. To overcome these limitations, two-dimensional (2D) layered graphene oxide (GO) has received much attention among the various 2D materials due to its ease of preparation as well as the tunability of its material properties [64 6]. The $Si_3N_4$ and Hydex platforms have been highly successful, forming the basis of integrated Kerr microcombs, largely a result of their negligible TPA and extremely high nonlinear FOM >>1 in the telecom band. For these platforms, however, their low intrinsic Kerr nonlinearity still poses a fundamental limitation with respect to the nonlinear efficiency [67, 68  9, 10]. Here, we demonstrate significantly improved SPM performance for $Si_3N_4$ waveguides integrated with 2D GO films. We perform SPM measurements using both picosecond and femtosecond optical pulses centered at telecom wavelengths. Compared to the uncoated $Si_3N_4$ waveguide, the GO-coated waveguides show more significant spectral broadening for both the picosecond and femtosecond optical pulses, achieving a maximum BF of ~3.4 for a device with 2 layers of GO. We also fit the SPM experimental results with theory and obtain a Kerr coefficient ($n_2$) for GO that is about 5 orders of magnitude higher than $Si_3N_4$. Finally, we discuss the influence of GO film's length and coating position on the SPM performance. For Si waveguides the main challenge is to enhance the nonlinear FOM, whereas for $Si_3N_4$ waveguides the challenge is to enhance the nonlinear parameter $\gamma$ ($= 2\pi n_2 / (\lambda A_{eff})$, where $A_{eff}$ is the effective mode area) since their FOM is very large already. We obtain an enhancement in $\gamma$ by a factor of up to ~18.4 for a $Si_3N_4$ waveguide coated with 2 layers of GO, compared to the uncoated waveguide, accompanied by only a modest increase in the linear loss of about 3 dB/cm per layer of GO and with no measurable decrease in the nonlinear FOM. These results confirm the high nonlinear optical performance of $Si_3N_4$ waveguides integrated with 2D GO films.

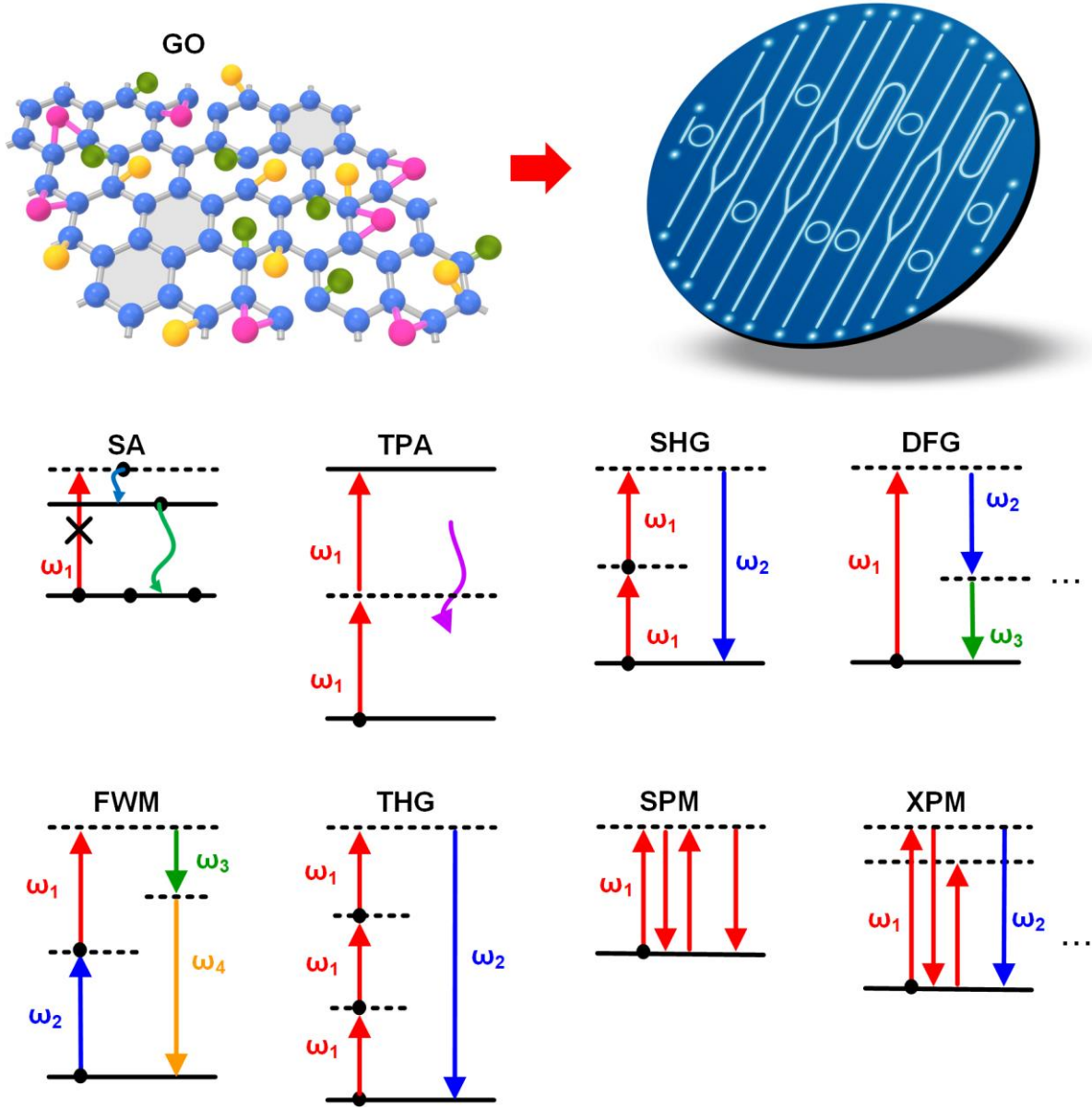

**Figure 1.** Schematic illustration of on-chip integration of GO for nonlinear optical applications. SA: saturable absorption. TPA: two-photon absorption. SHG: second-harmonic generation. DFG: difference frequency generation. FWM: four-wave mixing. THG: third-harmonic generation. SPM: self-phase modulation. XPM: cross-phase modulation.

## 2. SILICON NITRIDE DEVICE FABRICATION

Figure 2(a) shows a schematic of a GO-coated $Si_3N_4$ waveguide with a monolayer GO film. The bare $Si_3N_4$ waveguide has a cross-section of 1.60 μm × 0.72 μm, which was fabricated via a CMOS-compatible crack-free method [69-71]. First, two-step deposition of $Si_3N_4$ film (360-nm-thick layer in each step) was achieved via low-pressure chemical vapor deposition (LPCVD) for strain management and crack prevention. Next, 248-nm deep ultraviolet lithography and $CF_4/CH_2F_2/O_2$ fluorine-based dry etching were employed for patterning the $Si_3N_4$ waveguides. A silica upper cladding was then deposited using high-density plasma-enhanced chemical deposition (HDP-PECVD), followed by opening a window on it down to the top surface of the $Si_3N_4$ waveguides via lithography and dry etching processes. Finally, the 2D layered GO film was coated onto the $Si_3N_4$ waveguide by using a solution-based method that enabled transfer-free and layer-by-layer film coating, as reported previously [72-75]. Compared to the sophisticated film transfer processes employed for on-chip integration of other 2D materials such as graphene and TMDCs [76-78], our GO coating method is highly scalable, enabling precise control of the GO layer number (i.e., film thickness), large-area film coating, and good film attachment on integrated chips [72, 79]. Figures 2(b-i) and (b-ii) show a schematic cross section and the transverse electric (TE) mode profile of the GO-coated $Si_3N_4$ waveguide in Figure 2(a), respectively. The interaction between light and the GO film possessing an ultrahigh Kerr nonlinearity can be excited by the waveguide evanescent field, which underpins the enhancement of the SPM response in the hybrid waveguide.

Figure 2(c) shows a microscope image of a $Si_3N_4$ integrated chip uniformly coated with a monolayer GO film, where the coated GO film exhibits good morphology, high transmittance, and high uniformity. The opened window on the silica upper cladding of the uncoated $Si_3N_4$ chip enables control of the film length and placement of the GO film that are in contact with the $Si_3N_4$ waveguide. Note that this can also be realized by patterning GO films on planarized $Si_3N_4$ waveguides (without silica upper cladding) via lithography and lift-off processes, as we did in our previous work [80]. In this work, we used $Si_3N_4$ waveguides with opened windows mainly because they have lower coupling loss and propagation loss, which is beneficial for boosting the nonlinear response of SPM.

Figures 2(d-i) and (d-ii) show the measured Raman spectra of a $Si_3N_4$ chip before and after coating 2 layers of GO, respectively, where the presence of the representative D and G peaks of GO in the latter one verifies the successful on-chip integration of the GO film [73, 80 15, 22]. According to our previous measurements [74, 80, 81 16, 22, 23], the GO film thickness shows a near linear relationship with layer number at small film thickness (i.e., layer numbers < 100), and the thickness for 1 layer of GO is ~2.0 nm. For the GO-coated $Si_3N_4$ waveguides used in the following SPM measurements, the measured film thicknesses for 1 and 2 layers of GO are ~2.1 nm and ~ 4.3 nm, respectively.

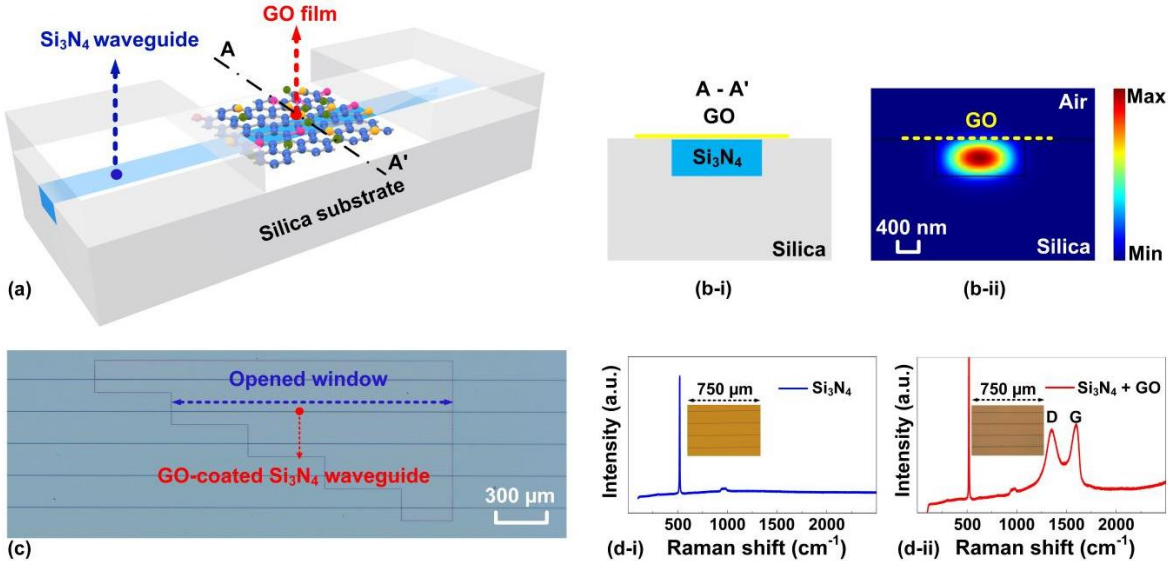

Figure 2. (a) Schematic illustration of a GO-coated $Si_3N_4$ waveguide with a monolayer GO film. (b-i) Schematic illustration of cross section and (b-ii) corresponding TE mode profile of the GO-coated $Si_3N_4$ waveguide in (a). (c) Microscope image of a $Si_3N_4$ integrated chip uniformly coated with monolayer GO film. (d) Raman spectra of a $Si_3N_4$ chip (i) before and (ii) after coating 2 layers of GO. Insets show the corresponding microscope images.

## 3. SILICON NITRIDE SPM MEASUREMENTS

Figure 3(a-i) shows the normalized spectra of picosecond optical pulses before and after propagation through the uncoated and GO-coated $Si_3N_4$ waveguides. The peak power of the input picosecond optical pulses was kept the same at ∼20 W. The output spectrum from the uncoated $Si_3N_4$ waveguide shows slight spectral broadening compared to the input pulse spectrum, which is mainly induced by the SPM in the $Si_3N_4$ waveguide. In contrast, the output spectra after propagating through the GO-coated $Si_3N_4$ waveguides show more significant spectral broadening, reflecting the enhanced SPM in these hybrid waveguides. Figure 3(a-ii) shows the output spectra after propagation through the hybrid waveguide with 2 layers of GO measured using picosecond optical pulses with different peak powers. We chose 6 different input peak powers ranging from 7 W to 20 W. As expected, the spectral broadening of the output spectra becomes more significant as the peak power increases. To quantitively compare the spectral broadening in these waveguides, we calculated the BFs for the measured output spectra. The BF is defined as [70, 81, 82]:

$$BF = \frac{\Delta\omega_{rms}}{\Delta\omega_0} \quad (1)$$

where $\Delta\omega_0$ and $\Delta\omega_{rms}$ are the root-mean-square (RMS) spectral widths of the input and output signals, respectively.

Figure 3(a-iii) shows the BFs for the uncoated and GO-coated $Si_3N_4$ waveguides versus the peak power of input picosecond optical pulses. As can be seen, the BFs for the GO-coated $Si_3N_4$ waveguides are higher than that of the uncoated waveguide, and the BF for the hybrid waveguide with 2 layers of GO is higher than that for the device with 1 layer of GO, showing agreement with the results in Figure 3(a-i). The BF increases with the peak power of the optical pulses, which is consistent with the results in Figure 3(a-ii). At a peak power of ~20 W, a maximum BF of ~1.3 is achieved for the hybrid waveguide with 2 layers of GO.

Figure 3(b-i) shows the normalized spectra of femtosecond optical pulses before and after propagation through the uncoated and GO-coated $Si_3N_4$ waveguides, which were measured at the same input peak power of ~160 W. Similar to Figure 3(a-i), the output spectra after passing through the hybrid waveguides show more significant spectral broadening compared to the uncoated waveguide. Figure 3(b-ii) shows the output spectra measured at different input peak powers for the hybrid waveguide with 2 layers of GO, showing the similar trend as that in Figure 3(a-ii). The peak power of the input femtosecond optical pulses ranged from 98 W to 160 W. The calculated BFs for the uncoated and GO-coated $Si_3N_4$ waveguides versus the input peak power are shown in Figure 3(b-iii). A maximum BF of ~3.4 is achieved at a peak power of 160 W for the hybrid waveguide with 2 layers of GO, which is ~2.6 times higher than the maximum BF achieved for the picosecond optical pulses. This mainly results from the relatively high peak power of the femtosecond optical pulses that drives more significant SPM in the hybrid waveguide.

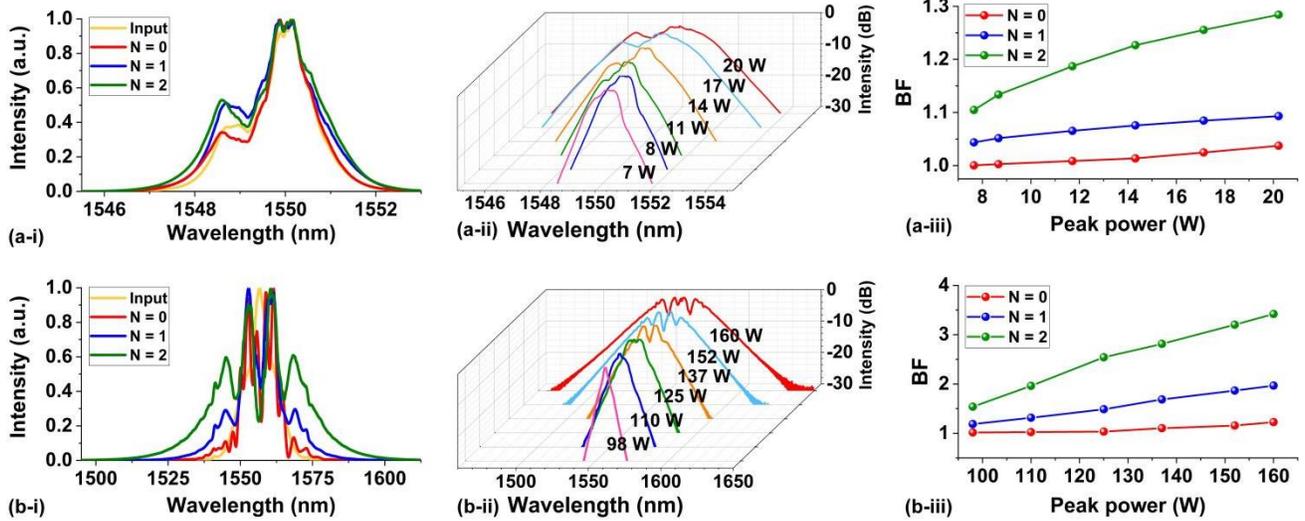

Figure 3. SPM experimental results using (a) picosecond optical pulses and (b) femtosecond optical pulses. (i) Normalized spectra of optical pulses before and after propagation through the GO-coated $Si_3N_4$ waveguides with 1 and 2 layers of GO. (ii) Optical spectra measured at different input peak powers for the hybrid waveguides with 2 layers of GO. (iii) BFs of the measured output spectra versus input peak power for the hybrid waveguides with 1 and 2 layers of GO. In (a) and (c), the corresponding results for the uncoated $Si_3N_4$ waveguides are also shown for comparison. SPM experimental results using femtosecond optical pulses.

## 4. SPM ANALYSIS AND DISCUSSION

Based on the theory in Refs. [64, 81, 83   8, 23, 25], we simulated the evolution of optical pulses traveling along the GO-coated $Si_3N_4$ waveguides using the nonlinear Schrodinger equation as follows:

$$\frac{\partial A}{\partial z} = -\frac{i\beta_2}{2}\frac{\partial^2 A}{\partial t^2} + i\gamma |A|^2 A - \frac{1}{2}\alpha A \quad (2)$$

where $i = \sqrt{1}$, $A(z, t)$ is the slowly varying temporal pulse envelope along the propagation direction $z$, $\beta_2$ is the second-order dispersion coefficient, and $\gamma$ is the waveguide nonlinear parameter. The overall loss factor $\alpha$ includes both the linear propagation loss and the SA-induced excess propagation loss.

Unlike in Refs. [81, 83], there are no free carrier absorption (FCA) and free carrier dispersion (FCD) items in Eq. (2) since the TPA in both $Si_3N_4$ and GO (with bandgaps > 2 eV) is negligible at near-infrared wavelengths. We retain only the second-order dispersion item in Eq. (2) because the physical length of the waveguides (20 mm) is much smaller than the dispersion length (> 1 m) [84]. In our simulation, we divided the GO-coated $Si_3N_4$ waveguides into uncoated (with silica cladding) and hybrid segments (coated with 1.4-mm-long GO films). Numerically solving Eq. (2) was performed for each segment, and the output from the previous segment was set as the input for the subsequent one.

Figures 4(a) and (b) show the spectrum evolutions for the picosecond and femtosecond optical pulses after transmission through the hybrid waveguides, respectively. (i) and (ii) show the hybrid waveguide with 1 and 2 layers of GO, respectively. As can be seen, the theoretical simulations also agree well with the experimental results. The fit $\gamma$'s for the hybrid waveguides with 1 and 2 layers of GO are ~11.5 and ~27.6, respectively, which are ~7.7 and ~18.4 times that of the uncoated $Si_3N_4$ waveguide, reflecting the significantly improved Kerr nonlinearity for the hybrid waveguides. The significant Kerr nonlinearity is also confirmed by the dramatical spectral broadening within the GO-coated region, as shown in the insets of Figures 4(a) and (b).

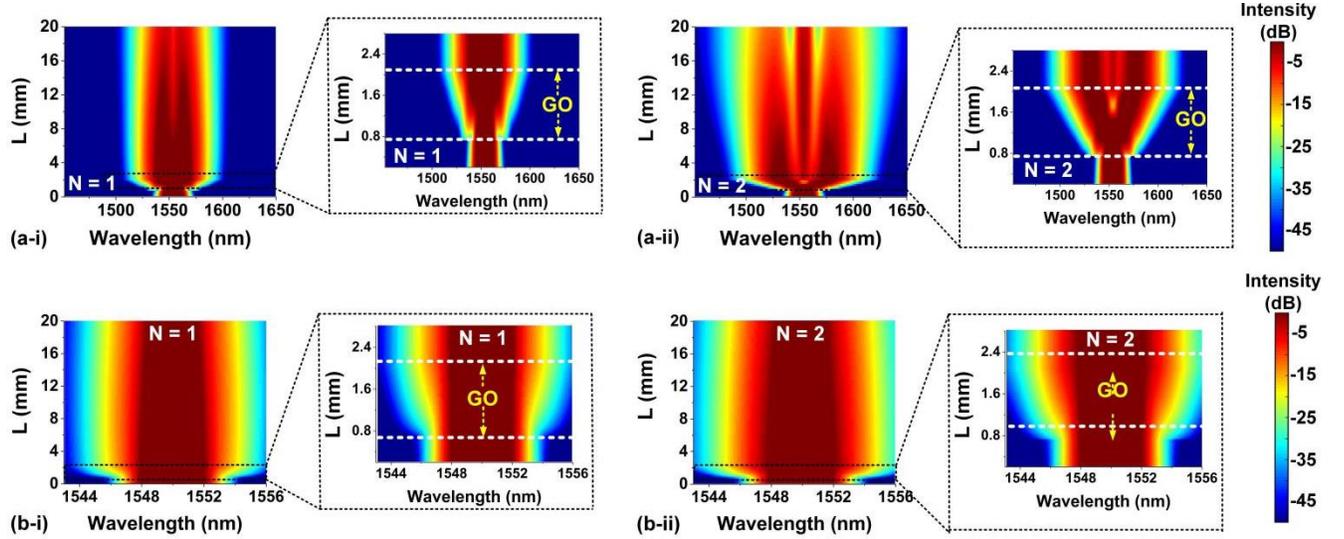

Figure 4. (a) Simulated spectra evolution of the picosecond optical pulses along the hybrid waveguide at the peak power ~20 W. (b) Simulated spectra evolution of the femtosecond optical pulses along the hybrid waveguide at the peak power ~160 W. Insets in (a) and (b) show zoom-in views for the GO-coated regions. (i) and (ii) show the hybrid waveguides with 1 layer and 2 layers of GO, respectively.

Based on the fit $\gamma$'s of the hybrid waveguides, we further extract the Kerr coefficient ($n_2$) of the layered GO films using [73, 85, 86] :

$$\gamma = \frac{2\pi}{\lambda_c} \frac{\iint_D n_0^2(x,y) n_2(x,y) S_z^2 dxdy}{\left[ \iint_D n_0(x,y) S_z dxdy \right]^2} \quad (3)$$

where $\lambda_c$ is the pulse central wavelength, $D$ is the integral of the optical fields over the material regions, $S_z$ is the time-averaged Poynting vector calculated using mode solving software, $n_0(x, y)$ is the refractive index profiles calculated over the waveguide cross section and $n_2(x, y)$ is the Kerr coefficient of the different material regions. The values of $n_2$ for silica and $Si_3N_4$ used in our calculation were $2.60 \times 10^{-20}$ $m^2$ $W^{-1}$ [87] and $2.59 \times 10^{-19}$ $m^2$ $W^{-1}$, respectively, with the latter obtained by fitting the experimental results for the uncoated $Si_3N_4$ waveguide.

The fit $\gamma$'s of the hybrid waveguides with 1 and 2 layers of GO are ~11.5 $W^{-1}m^{-1}$ and ~27.6 $W^{-1}m^{-1}$, respectively, which are ~7.7 and ~18.4 times that of the uncoated $Si_3N_4$ waveguide. The extracted $n_2$ of 1 and 2 layers of GO are ~$1.23 \times 10^{-14}$ $m^2$ $W^{-1}$ and ~$1.19 \times 10^{-14}$ $m^2$ $W^{-1}$, respectively. Both of the values are about 5 orders of magnitude higher than that of $Si_3N_4$ and agree reasonably well with our previous measurements [80]. Note that the $n_2$ of 1 layer of GO is higher than that of 2 layers of GO. We infer this may result from the increased inhomogeneous defects within the GO layers and imperfect contact between the multiple GO layers. Nonetheless, the higher GO mode overlap for the thicker 2-layer film, compared to the single-layer film, resulted in a more than doubling of the nonlinear parameter $\gamma$.

Based on the SPM modeling in Eq. (2) and the fit parameters obtained from Figure 4, we further investigate the influence of GO film length ($L_c$) and coating position ($L_0$) on the SPM performance of GO-coated $Si_3N_4$ waveguides.

Figures 5(a) and (b) show the calculated BFs versus $L_c$ and input peak power ($P_{peak}$) for picosecond and femtosecond optical pulses after propagation through the hybrid waveguides, respectively. In each figure, (i) and (ii) show the results for the waveguides with 1 and 2 layers of GO, respectively. The coating position is fixed at $L_0 = 0.7$ mm. The black points mark the parameters corresponding to the SPM measurements, where the calculated BFs are consistent with the experimental results in Figure 3. The BF increases with both $L_c$ and $P_{peak}$, with maximum BFs of 4.2 (at $L_c = 19.3$ mm and $P_{peak} = 30$ W) and 25.0 (at $L_c = 19.3$ mm and $P_{peak} = 240$ W) being achieved for the picosecond and femtosecond optical pulses, respectively. This reflects that there is a large room for improvement in the SPM-induced spectral broadening by increasing the GO film length and the input peak power. The BF can also be improved by coating thicker GO films (i.e., $N > 2$), which was used for increasing the FWM conversion efficiency in [80]. The increased GO film thickness will also lead to loss increase for the hybrid waveguides and hence creates a need to balance the trade-off between the Kerr nonlinearity and loss [67, 68].

Figures 5(c) and (d) show the calculated BFs versus $L_0$ and $P_{peak}$ for picosecond and femtosecond optical pulses after propagation through the hybrid waveguides, respectively, where the film length is fixed at $L_c = 1.4$ mm. The simulation results marked by the black points also agree well with the experimental results in Figure 3. The BF increases with $P_{peak}$ – a trend similar to that in Figures 5(a) and (b). In contrast, it decreases with $L_0$, with the maximum value being achieved at $L_0 = 0$. This indicates that the largest spectral broadening can be achieved by coating GO films at the beginning, as expected since the light power is highest at the start of the waveguide.

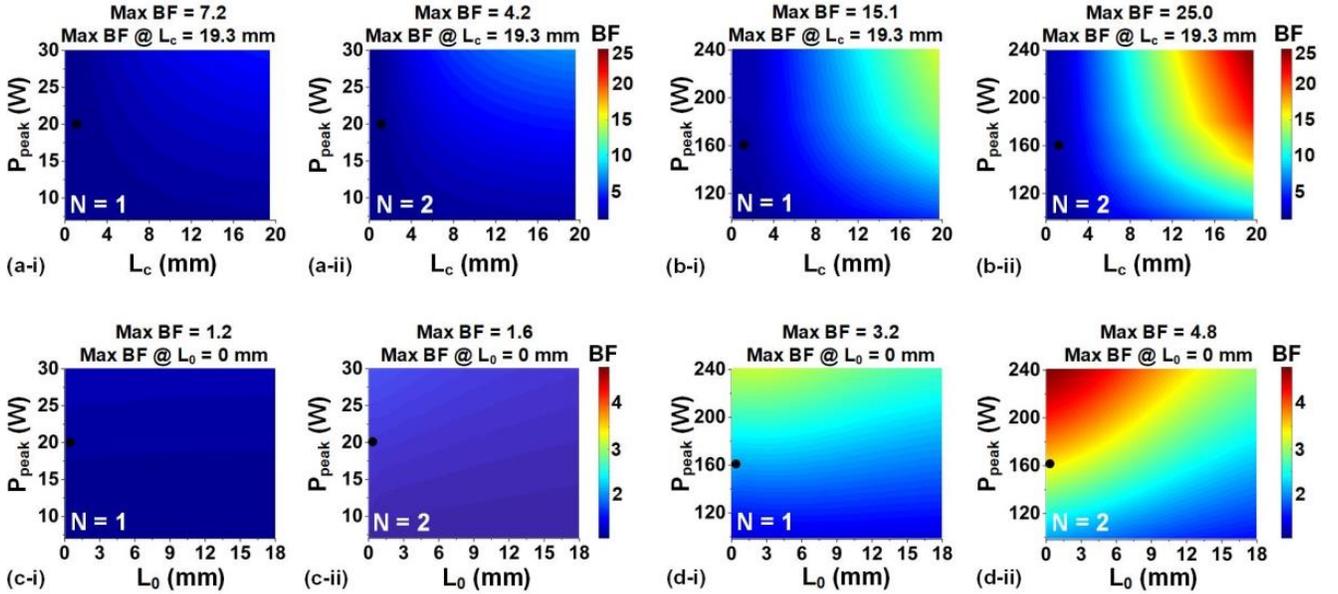

Figure 5. (a) BFs versus GO film length ($L_c$) and input peak power ($P_{peak}$) for picosecond optical pulses after propagation through the hybrid waveguides. (b) BFs versus $L_c$ and $P_{peak}$ for femtosecond optical pulses after propagation through the hybrid waveguides. (c) BFs versus GO coating position ($L_0$) and $P_{peak}$ for picosecond optical pulses after propagation through the hybrid waveguides. (d) BFs versus $L_0$ and $P_{peak}$ for femtosecond optical pulses after propagation through the hybrid waveguides. In (a) – (d), (i) and (ii) show the corresponding results for the waveguides with 1 and 2 layers of GO, the black points mark the results corresponding to the device parameters and input powers in Figures 3. In (a) and (b), $L_0 = 0.7$ mm. In (c) and (d), $L_c = 1.4$ mm.

In Table 1, we provide comparisons for the parameters of the GO-$Si_3N_4$ waveguides in this work and the GO-Si waveguides in Ref. [81]. We note that the trade-offs and challenges involved with integrating GO films into these two very different platforms, are in turn very different. Compared to the GO-Si waveguides, GO-$Si_3N_4$ waveguides have a larger waveguide geometry, which results in lower mode overlap with GO films. Such a reduced GO mode overlap yields a lower GO-induced excess propagation loss, at the expense of a weaker light-GO interaction. Despite this, the nonlinear parameter $\gamma$ of the GO-$Si_3N_4$ waveguide with 1 layer of GO is still ~7.7 times that of the uncoated waveguide. In contrast, there is only about a 2-fold improvement in the $\gamma$ of the GO-Si waveguide with 1 layer of GO. This mainly due to the relatively low $n_2$ of $Si_3N_4$ compared to Si, reflecting that integrating GO onto $Si_3N_4$ waveguides has a more dramatic impact on improving the nonlinear performance. In contrast to Si that has strong TPA at near infrared wavelengths, the TPA of $Si_3N_4$ in this wavelength range is absent, which yields much higher values of nonlinear FOM for both the uncoated and GO-coated $Si_3N_4$ waveguides. Hence, the motivation in integrating GO films onto Si waveguides lies very much in increasing the nonlinear FOM, whereas for Si$_3$N$_4$ waveguides, the main benefit of integrating GO films is to increase the nonlinearity (i.e., nonlinear parameter $\gamma$) without introducing additional nonlinear loss. Potentially this could be useful for enhancing nonlinear optics for applications to optical microcombs. [88-100]

Table 1. Performance comparison of SOI nanowire and SiN waveguides integrated with 2D layered GO films.

| Parameters | Refractive index [a] | EPL$_{GO-1}$ [b] (dB/cm) | $\gamma_{WG}$ [c] (W$^{-1}$m$^{-1}$) | $\gamma_{hybrid}$ [d] (W$^{-1}$m$^{-1}$) | Fit $n_2$ of GO ($\times 10^{-14}$ m$^2$/W) | FOM [e] | Ref. |
|---|---|---|---|---|---|---|---|
| GO-Si | 3.48 | 20.5 | 288.0 | 668.0 ($N = 1$) | 1.42 ($N = 1$) | 1.1 ($N = 1$) | [81] |
| | | | | 990.0 ($N = 2$) | 1.33 ($N = 2$) | 2.4 ($N = 2$) | |
| GO-Si$_3$N$_4$ | 1.99 | 3.0 | 1.5 | 11.5 ($N = 1$) | 1.23 ($N = 1$) | $\gg 1$ | This work |
| | | | | 27.6 ($N = 2$) | 1.19 ($N = 2$) | | |

[a] These values are at 1550 nm.
[b] EPL$_{GO-1}$: excess propagation loss induced by GO for the hybrid waveguide with 1 layer of GO.
[c] $\gamma_{WG}$: nonlinear parameters of the bare waveguides.
[d] $\gamma_{hybrid}$: nonlinear parameter of the hybrid waveguide with 1 layer (GO-1) and 10 (GO-2) layers of GO.
[e] The definition of $FOM = n_2 / (\lambda \beta_{TPA})$ is the same as those in Refs. [83, 85], with $n_2$ and $\beta_{TPA}$ denoting the effective Kerr coefficient and TPA coefficient of the waveguides, respectively, and $\lambda$ the light wavelength at 1550 nm.

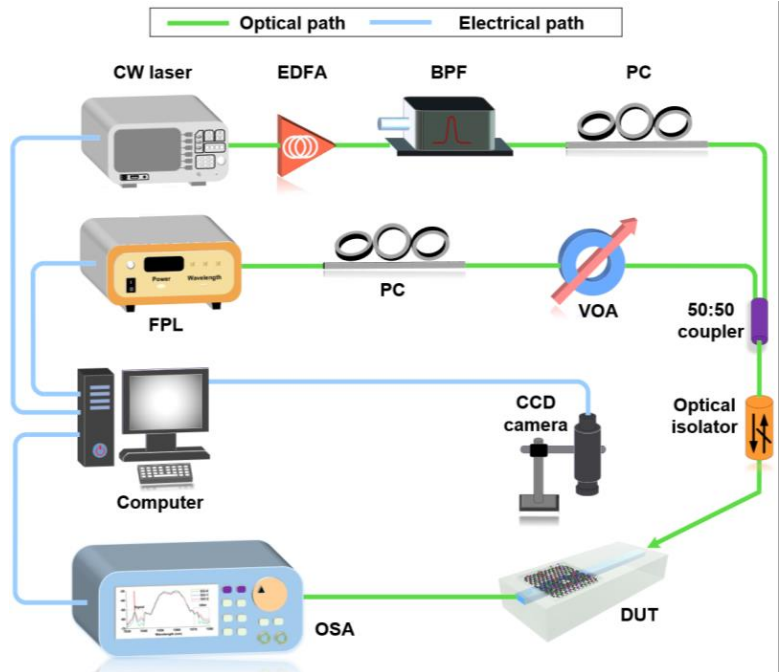

**Figure 6.** Experimental setup for OPA experiments. CW laser: continuous-wave laser. FPL: fiber pulsed laser. PC: polarization controller. EDFA: Erbium doped fiber amplifier. VOA: variable optical attenuator. OPM: optical power meter. DUT: device under test. CCD: charged-coupled device. OSA: optical spectrum analyzer.

## 5. OPTICAL PARAMETRIC AMPLIFICATION

Optical amplifiers are key to many applications [101-103] such as optical communications where they have been instrumental with rare-earth-doped fibers [104-106] and III-V semiconductors [107-109]. However, these devices are restricted to specific wavelength ranges determined by the energy gaps between states [101-110]. In contrast, optical parametric amplification (OPA) can achieve gain across virtually any wavelength range [111-112], and so is capable of achieving broadband optical amplification outside of conventional wavelength windows [111-113]. Since its discovery in 1965[114], OPA has found applications in many fields such as ultrafast spectroscopy[115-116], optical communications[105-113], optical imaging[117-118], laser processing[119-120], and quantum optics[121-122]. Notably, it has underpinned many new technological breakthroughs such as optical microcombs[123-124] and entangled photon pairs[125-126].

To achieve OPA, materials with a high optical nonlinearity are needed – either second- ($\chi^{(2)}$) or third-order ($\chi^{(3)}$) nonlinearities[127-128], and has been demonstrated in birefringent crystals[129-131], optical fibers [110, 132, 133], and photonic integrated chips[101, 103, 124, 134, 135]. Amongst these, photonic integrated chips offer the advantages of a compact footprint, low power consumption, high stability and scalability, as well as cost reduction through large-scale manufacturing[136-138]. Despite silicon's dominance as a platform for linear photonic integrated devices[139,140], its significant two photon absorption (TPA) in the near infrared wavelength region and the resulting free carrier absorption lead to a high nonlinear loss [103, 127], making it challenging to achieve any significant OPA gain in this wavelength range. Other nonlinear integrated material platforms, such as silicon nitride ($Si_3N_4$) [101, 141], silicon rich nitride [142, 143], doped silica [136, 144], AlGaAs[145-146], chalcogenide [147, 148], GaP [149], and tantala [150], exhibit much lower TPA at near infrared wavelengths and have made significant progress over the past decade. However, their comparatively low third-order optical nonlinearity imposes a significant limitation on the OPA gain that they can achieve.

Recently, two-dimensional (2D) materials with ultrahigh optical nonlinearities and broadband response have been integrated on photonic chips to achieve exceptional nonlinear optical performance [125,151-154], highlighted by the progress in realizing OPA by exploiting the high second-order optical nonlinearities of monolayer transition metal dichalcogenides (TMDCs) [125]. Previously[155-159], we reported an ultra-high third-order optical nonlinearity in 2D graphene oxide (GO) films that is about 4 orders of magnitude larger than silicon, together with a large bandgap (> 2 eV) that yields a linear loss more than 2 orders of magnitude lower than graphene, and perhaps most importantly, low TPA at near infrared wavelengths – all of which are key to achieving high OPA. In addition, GO has demonstrated high compatibility with various integrated platforms [112, 138], along with the capability to achieve precise control over its film thickness and length [156, 160].

In this work, we demonstrate significantly increased optical parametric gain in $Si_3N_4$ waveguides by integrating them with 2D layered GO films. We employ a transfer-free, layer-by-layer coating method to achieve precise control over the GO film thickness, and by using photolithography to open windows in the waveguide cladding we are able to accurately control the GO film length and position. We perform a detailed experimental characterization of the OPA performance of the devices with different GO film thicknesses and lengths, achieving a maximum parametric gain of ~24.0 dB, representing a ~12.2 dB improvement over the uncoated device. By fitting experimental results with theory, we analyse the influence of the applied power, wavelength detuning, and GO film thickness and length on the OPA performance, and in the process demonstrate that there is still significant potential for improved performance. These results verify the effectiveness of the on-chip integration of 2D GO films to improve the OPA performance of photonic integrated devices.

## 6. OPTICAL PARAMETRIC AMPLIFICATION EXPERIMENTS

We conducted OPA experiments using the same devices that were fabricated and used for the loss measurements. A schematic of the experimental setup is shown in **Figure 6**. To generate the pump light required for the OPA experiments, we employed the same FPL that was used for the loss measurements. On the other hand, the signal light was generated through amplification of the CW light from a tunable laser. The pulsed pump and the CW signal were combined by a broadband 50:50 coupler and sent to the device under test (DUT) for the optical parametric process. The polarization of both signals was adjusted to TE polarized using two polarization controllers (PCs). To adjust the power of the pulsed pump, a broadband variable optical attenuator (VOA) was utilized. The output after propagation through the DUT was directed towards an optical spectrum analyzer (OSA) for analysis.

**Figure 7a** shows the optical spectra after propagation through the uncoated $Si_3N_4$ waveguide and the hybrid waveguides with 1 and 2 layers of GO. For all three devices, the input pump peak power and signal power were kept the same at $P_{peak}$ = ~180 W and $P_{signal}$ = ~6 mW, respectively. As the pump light used for the OPA experiments was pulsed, the optical parametric process occurred at a rate equivalent to the repetition rate of the FPL. As a result, both the generated idler and amplified signal also exhibited a pulsed nature with the same repetition rate as that of the FPL. The optical spectra in **Figure 7a** were analyzed to extract the parametric gain $PG$ experienced by the signal light for the three devices. The $PG$ for the uncoated $Si_3N_4$ waveguide and the hybrid waveguides with 1 and 2 layers of GO were ~11.8 dB, ~20.4 dB, and ~24.0 dB, respectively. The hybrid waveguides exhibited higher parametric gain compared to the uncoated waveguide, and the 2-layer device had higher parametric gain than the 1-layer device. These results confirm the improved OPA performance in the $Si_3N_4$ waveguide by integrating it with 2D GO films. We also note that the hybrid devices showed greater spectral broadening of the pulsed pump caused by self-phase modulation (SPM), which is consistent with our previous observations from SPM experiments [157].

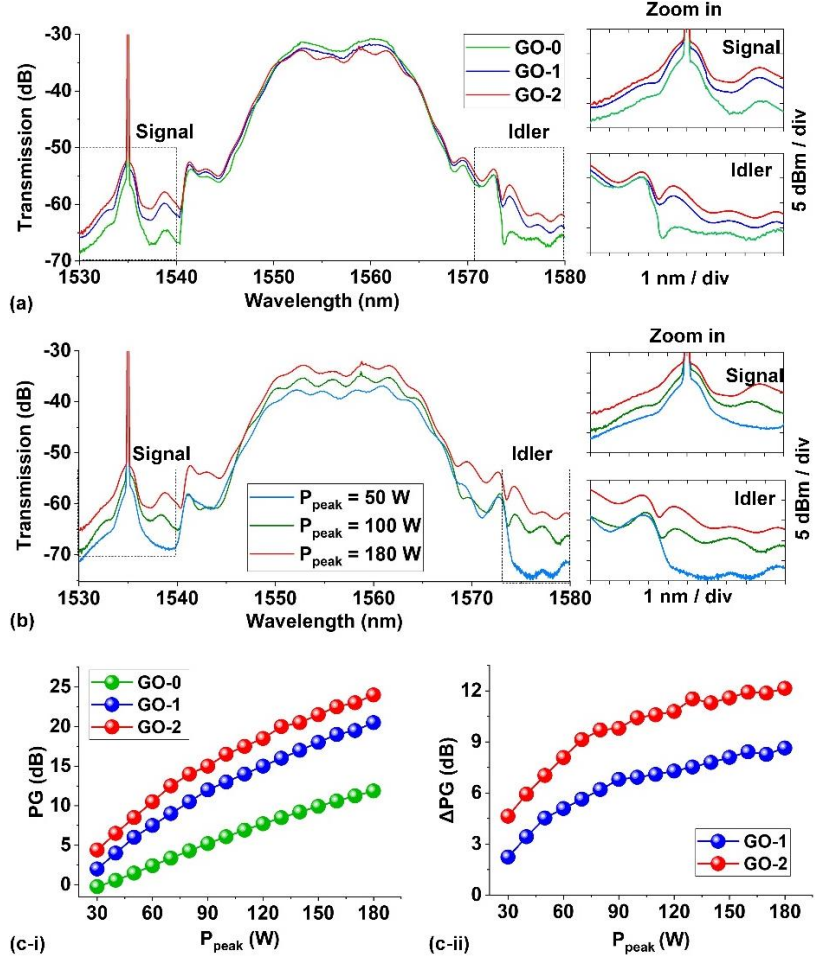

**Figure 7.** Optical parametric amplification (OPA) using a 180-fs pulsed pump and a continuous-wave (CW) signal. (a) Measured output optical spectra after propagation through uncoated (GO-0) and hybrid waveguides with 1 (GO-1) and 2 (GO-2) layers of GO. The peak power of the input pump light $P_{peak}$ was ~180 W. (b) Measured output optical spectra after propagation through the device with 2 layers of GO at different $P_{peak}$. In (a) and (b), the power of the CW signal light was $P_{signal}$ = ~6 mW, and insets show zoom-in views around the signal and idler. (c) Measured (i) parametric gain $PG$ and (ii) parametric gain improvement $\Delta PG$ versus $P_{peak}$.

The values of $PG$ in **Figure 7** are the net parametric gain, over and above the waveguide loss induced by both the GO-coated and uncoated $Si_3N_4$ waveguide segments. This is different to the "on/off" parametric gain often quoted [111, 143], where the waveguide loss is excluded, resulting in higher values of parametric gain. Here, the on-off gains for the waveguides with 0, 1, and 2 layers of GO were ~13. 2 dB, ~22.3 dB, and ~26.2 dB, respectively, which are only slightly higher than their corresponding net gains due to the low loss of the $Si_3N_4$ waveguides and the relatively short GO film length. Although the net gain can be increased closer to the on-off gain by reducing the waveguide loss via optimization of the fabrication processes, because the differences between the net and on-off gains are small in our case, there is not much incentive to do this. In the following, we focus our discussion on the net parametric gain $PG$. This can also ensure a fair comparison of the parametric gain improvement, as different waveguides have different waveguide loss.

**Figure 7b** shows the measured output optical spectra after propagation through the device with 2 layers of GO for different $P_{peak}$. **Figure 7c-i** shows the signal parametric gain $PG$ for the uncoated and hybrid waveguides versus input pump peak power, and the parametric gain improvement $\Delta PG$ for the hybrid waveguides as compared to the uncoated waveguide is further extracted and shown in **Figure 7c-ii**. We varied the input pump peak power from ~30 W to ~180 W, which corresponds to the same power range used for loss measurements. The $PG$ is higher for the hybrid waveguide with 1 layer of GO compared to the uncoated waveguide, and lower than the device with 2 layers of GO. In addition, both $PG$ and $\Delta PG$ increase with $P_{peak}$, and a maximum $\Delta PG$ of ~12.2 dB was achieved for the 2-layer device at $P_{peak}$ = ~180 W. Likewise, we observed similar phenomena when using lower-peak-power picosecond optical pulses for the pump.

To evaluate the OPA performance, we conducted experiments where we varied the wavelength detuning, CW signal power, and GO film length. Except for the varied parameters, all other parameters are the same as those in **Figure 7**. In **Figure 8a**, the measured signal parametric gain *PG* and parametric gain improvement Δ*PG* are plotted against the wavelength detuning Δλ, which is defined as the difference between the CW signal wavelength $λ_{signal}$ and the pump center wavelength $λ_{pump}$. It is observed that both the *PG* and Δ*PG* increase as Δλ changes from -12 nm to -22 nm. In **Figure 8b**, the *PG* and Δ*PG* are plotted against the CW signal power $P_{signal}$, showing a slight decrease as $P_{signal}$ increases, which is primarily due to the fact that an increase in $P_{signal}$ can result in a decrease in *PG* as per its definition (i.e., *PG* = $P_{out,signal}$ / $P_{in,signal}$). **Figure 8c** shows the *PG* and Δ*PG* versus GO film length. By measuring devices with various GO film lengths, ranging from ~0.2 mm to ~1.4 mm, we observed that those with longer GO films exhibited greater *PG* and Δ*PG* values. The *PG* achieved through the optical parametric process is influenced by several factors, such as the applied powers, optical nonlinearity, dispersion, and loss of the waveguides. These factors will be comprehensively analyzed in the following section.

## 7. OPA ANALYSIS AND DISCUSSION

**Optical nonlinearity of hybrid waveguides and GO films.** We used the theory from Refs. [110, 158,170] to model the OPA process in the fabricated devices. By fitting the measured *PG* with theory, we obtained the nonlinear parameter γ of the uncoated and hybrid waveguides. The fit γ for the uncoated $Si_3N_4$ waveguide is ~1.11 $W^{-1}m^{-1}$, which is consistent with the previously reported values in the literature [158, 171 - 182]. **Figure 9a** shows the fit γ of the hybrid waveguides as a function of pulse peak power $P_{peak}$. For both devices with different GO film thickness, the lack of any significant variation in γ with $P_{peak}$ indicates that the applied power has a negligible effect on the properties of the GO films. This is in contrast to the effects of light with high average optical powers, which can lead to changes in GO's properties via photo-thermal reduction[156, 158]. The fit values of γ for the devices with 1 and 2 layers of GO are ~14.5 and ~27.3 times greater than the value for the uncoated $Si_3N_4$ waveguide. These agree with our earlier work [158, 159] and indicate a significant improvement in Kerr nonlinearity for the hybrid waveguides.

Based on the fit γ for the hybrid waveguides, we further extracted the Kerr coefficient $n_2$ of the GO films, as shown in **Figure 9b**. The extracted $n_2$ values for the films with 1 and 2 layers are similar, with the former being slightly higher than the latter. The lower $n_2$ for thicker films is likely caused by an increase in inhomogeneous defects within the GO layers and imperfect contact between multiple GO layers. The $n_2$ values for the films with 1 and 2 layers are about 5 orders of magnitude higher than that of $Si_3N_4$ (~2.62 × $10^{-19}$ $m^2$/W, obtained by fitting the result for the uncoated $Si_3N_4$ waveguide), highlighting the tremendous third-order optical nonlinearity of the GO films.

We also quantitatively compare the nonlinear optical performance of the $Si_3N_4$ waveguide and the hybrid waveguides by calculating their nonlinear figure of merit *FOM*. The *FOM* is determined by balancing a waveguide's nonlinear parameter against its linear propagation loss, and can be expressed as a function of waveguide length *L* given by:

$$FOM (L) = γ × L_{eff} (L) \qquad (4)$$

where γ is the waveguide nonlinear parameter and $L_{eff}(L) = [1 - exp(-α×L)]/α$ is the effective interaction length, with α denoting the linear loss attenuation coefficient. Note that the nonlinear figure of merit defined in **Eq. (4)** allows for comparison of the nonlinear optical performance of optical waveguides made from different materials. This is distinct from the nonlinear figure of merit commonly used for comparing the nonlinear optical performance of a single material, which is defined as $n_2/(λ·β_{TPA})$ [136], with $n_2$, λ, and $β_{TPA}$ denoting the Kerr coefficient, wavelength, TPA coefficient, respectively.

**Figure 9c** shows $L_{eff}$ versus *L* for the $Si_3N_4$ waveguide and the hybrid waveguides with 1 and 2 layers of GO. The $Si_3N_4$ waveguide has a higher $L_{eff}$ due to its comparably lower linear propagation loss. **Figure 9d** shows the *FOM* versus *L* for the three waveguides. Despite having a lower $L_{eff}$, the hybrid waveguides exhibit a higher *FOM* than the $Si_3N_4$ waveguide, owing to the significantly improved nonlinear parameter γ for the hybrid waveguides. This indicates that the impact of enhancing the optical nonlinearity is much greater than the degradation caused by the increase in loss, resulting in a significant improvement in the device's overall nonlinear optical performance.

For the hybrid waveguides that we measured in the OPA experiments, only a specific section of the waveguides was coated with GO films. In **Figures 9e** and **9f**, we compare *PG* and Δ*PG* versus waveguide length *L* for the hybrid waveguides uniformly coated with GO films, respectively, which were calculated based on the fit γ values (at $P_{peak}$ = ~180 W) in **Figure 9a**. The pump peak power, CW signal power, and wavelength detuning were $P_{peak}$ = ~180 W, $P_{signal}$ = ~6 mW, and Δλ = ~-22 nm, respectively – the same as those in **Figure 7a**. The corresponding results for the uncoated $Si_3N_4$ waveguide are also shown for comparison. The 2-layer device has higher *PG* and Δ*PG* values for *L* < ~5.7 mm but lower values for *L* > ~5.7 mm, reflecting the trade-off between the increase in optical nonlinearity and waveguide loss. At *L* = 1.4 mm, the 1-layer and 2-layer devices achieve *PG* of ~10.5 dB and ~15.6 dB, respectively. When compared to waveguides that have patterned GO films of the same length as those used in our OPA experiments, their total *PG* (including those provided by both the ~1.4-mm-long GO-coated section and the ~18.6-mm-long uncoated section) are ~20.4 dB and ~24.0 dB, respectively. This highlights the dominant role of the GO-coated section in providing the parametric gain, as well as the fact that a further improvement in Δ*PG* could be obtained by increasing the length of the GO-coated segments.

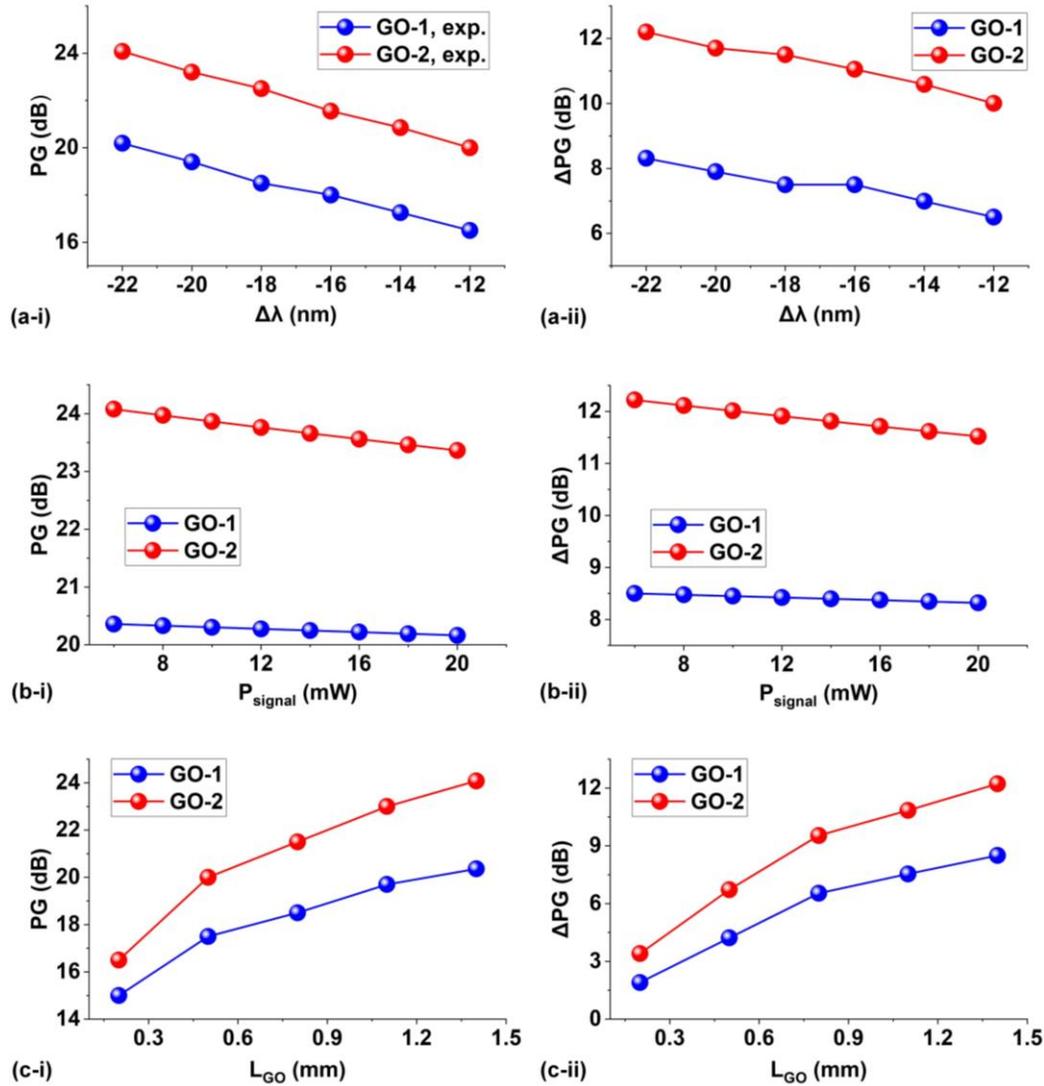

**Figure 8.** (a) Measured (i) parametric gain *PG* and (ii) parametric gain improvement Δ*PG* versus wavelength detuning Δ*λ*. (b) Measured (i) *PG* and (ii) Δ*PG* versus input CW signal power $P_{signal}$. (c) Measured (i) *PG* and (ii) Δ*PG* versus GO film length $L_{GO}$. In (a) – (c), the peak power of the 180-fs pulsed pump centered around 1557 nm was $P_{peak}$ = ~180 W. Except for the varied parameters, all other parameters are kept the same as Δ*λ* = ~-22 nm, $P_{signal}$ = ~6 mW, and $L_{GO}$ = ~1.4 mm.

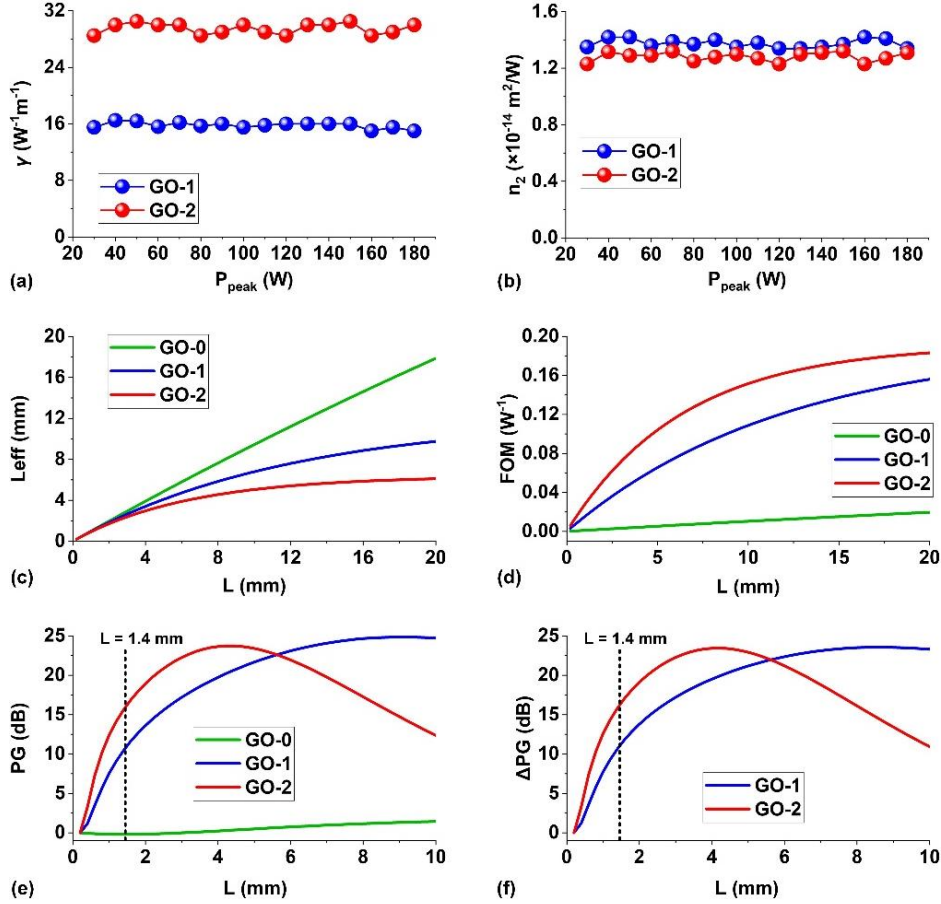

**Figure 9.** (a) Nonlinear parameter $\gamma$ of hybrid waveguides with 1 (GO-1) and 2 (GO-2) layers of GO as a function of pump peak power $P_{peak}$. (b) Kerr coefficient $n_2$ of films with 1 (GO-1) and 2 (GO-2) layers of GO versus $P_{peak}$. (c) Effective interaction length $L_{eff}$ and (d) figure of merit $FOM$ versus waveguide length $L$ for the uncoated (GO-0) and hybrid waveguides with 1 (GO-1) and 2 (GO-2) layers of GO. (e) Parametric gain $PG$ and (f) parametric gain improvement $\Delta PG$ versus waveguide length $L$ for the uncoated $Si_3N_4$ waveguide (GO-0) and the hybrid waveguides uniformly coated with 1 (GO-1) and 2 (GO-2) layers of GO. In (e) and (f), the pump peak power, CW signal power, and the wavelength detuning are $P_{peak} = \sim180$ W, $P_{signal} = \sim6$ mW, and $\Delta\lambda = \sim-22$ nm, respectively.

## 8. CONCLUSION

The on-chip integration of GO with a large optical nonlinearity and a high degree of flexibility in changing its properties represents a promising frontier for implementing high-performance nonlinear integrated photonic devices for a wide range of applications. In this paper, we present our recent progress in GO nonlinear integrated photonics. We summarize the optical properties of GO and the fabrication technologies for its on-chip integration. We review a range of GO hybrid integrated devices for different nonlinear optical applications, and compare the nonlinear optical performance of different integrated platforms. We also discuss the challenges and perspectives of this nascent field. Accompanying the advances in this interdisciplinary field, we believe that GO based nonlinear integrated photonics will become a new paradigm for both scientific research and industrial applications in exploiting the enormous opportunities arising from the merging of integrated devices and 2D materials.